\documentclass[sigconf, language=french, language=english]{acmart}
\usepackage{textcomp}
\let\oldftcp\footnotetextcopyrightpermission
\renewcommand\footnotetextcopyrightpermission[1]{\oldftcp{%
    \textcopyright{} 2023, Copyright is with the authors.
    Published in the Proceedings of the BDA 2023 Conference
    (October 23-26, 2023, Montpellier, France).
    Distribution of this article is permitted under the terms
    of the Creative Commons license CC-by-nc-nd 4.0.\\
    \textcopyright{} 2023, Droits restant aux auteurs.
    Publi{\'e} dans les actes de la conf{\'e}rence BDA 2023
    (23-26 octobre 2023, Montpellier, France).
    Redistribution de cet article autoris{\'e}e selon les termes
    de la licence Creative Commons CC-by-nc-nd 4.0.}}
\usepackage{listings}

\AtBeginDocument{%
  }

\setcopyright{none}
\copyrightyear{2023}
\acmYear{2023}
\acmDOI{}

\acmConference[BDA]{Bases de Données Avancées}{23-26 oct. 2023}{Montpellier, FRANCE}
\acmBooktitle{BDA 2023}
\acmPrice{}
\acmISBN{}

\begin{document}

%%
%% The "title" command has an optional parameter,
%% allowing the author to define a "short title" to be used in page headers.
\title{Graph versioning for evolving urban data}
\translatedtitle{french}{Versionnement de graphe pour les données urbaines évolutives}

%%
%% The "author" command and its associated commands are used to define
%% the authors and their affiliations.
%% Of note is the shared affiliation of the first two authors, and the
%% "authornote" and "authornotemark" commands
%% used to denote shared contribution to the research.
\author{Jey Puget Gil}
\email{jey.puget-gil@liris.cnrs.fr}
\orcid{0009-0006-6198-7488}
\affiliation{%
  \institution{Université de Lyon, Université Claude Bernard, LIRIS, UMR-CNRS 5205}
  \streetaddress{25 avenue Pierre de Coubertin}
  \city{Villeurbanne}
  \country{FRANCE}
  \postcode{69622 Cedex}
}

\author{Emmanuel Coquery}
\email{emmanuel.coquery@univ-lyon1.fr}
\orcid{}
\affiliation{%
  \institution{Université Claude Bernard, LIRIS, UMR-CNRS 5205}
  \streetaddress{25 avenue Pierre de Coubertin}
  \city{Villeurbanne}
  \country{FRANCE}
  \postcode{69622 Cedex}
}

\author{John Samuel}
\email{john.samuel@cpe.fr}
\orcid{0000-0001-8721-7007}
\affiliation{%
  \institution{Université de Lyon, CPE Lyon, LIRIS, UMR-CNRS 5205}
  \streetaddress{43, boulevard du 11 Novembre 1918}
  \city{Villeurbanne}
  \country{FRANCE}
  \postcode{69616 Cedex}
}

\author{Gilles Gesquière}
\email{gilles.gesquiere@univ-lyon2.fr}
\orcid{0000-0001-7088-1067}
\affiliation{%
  \institution{Université de Lyon, Université Lumière Lyon 2, LIRIS, UMR-CNRS 5205}
  \streetaddress{25 avenue Pierre de Coubertin}
  \city{Villeurbanne}
  \country{FRANCE}
  \postcode{69622 Cedex}
}

%%
%% By default, the full list of authors will be used in the page
%% headers. Often, this list is too long, and will overlap
%% other information printed in the page headers. This command allows
%% the author to define a more concise list
%% of authors' names for this purpose.
% \renewcommand{\shortauthors}{Trovato et al.}

%%
%% The abstract is a short summary of the work to be presented in the
%% article.
\begin{abstract}
The continuous evolution of cities poses significant challenges in terms of managing and understanding their complex dynamics.
With the increasing demand for transparency and the growing availability of open urban data,
it has become important to ensure the reproducibility of scientific research and computations in urban planning.
To understand past decisions and other possible scenarios, we require solutions that go beyond the management of urban knowledge graphs.
In this work, we explore existing solutions and their limits and explain the need and possible approaches for querying across multiple graph versions.
\end{abstract}

\begin{translatedabstract}{french}
L'évolution continue des villes pose des défis importants en termes de gestion et de compréhension de leurs dynamiques complexes.
Avec la demande croissante de transparence et la disponibilité grandissante de données urbaines ouvertes, il est devenu important d'assurer la reproductibilité de la recherche scientifique et des calculs dans le domaine de l'urbanisme.
Pour comprendre les décisions passées et d'autres scénarios possibles, nous avons besoin de solutions qui vont au-delà de la gestion des graphes de connaissances urbaines.
Dans ce travail, nous explorons les solutions existantes et leurs limites, et expliquons le besoin et les approches possibles pour l'interrogation à travers de multiples versions de graphes.
\end{translatedabstract}

%%
%% The code below is generated by the tool at http://dl.acm.org/ccs.cfm.
%% Please copy and paste the code instead of the example below.
%%


%%
%% Keywords. The author(s) should pick words that accurately describe
%% the work being presented. Separate the keywords with commas.
\translatedkeywords{french}{RDF, versionnement, graphe, données urbaines, déduction}

\received{June 7, 2023}
\received[revised]{August 21, 2023}
% \received[accepted]{}

%%
%% This command processes the author and affiliation and title
%% information and builds the first part of the formatted document.
\maketitle

\section{Introduction and Motivation}
Urban planners, historians, archaeologists, and researchers are continuously analyzing the constant development of cities.
They are interested in an understanding of the possible versions and scenarios of the city \cite{samuel_representation_2020}, both in the past and in the future, if certain decisions were to be made.
The choices made and the lessons learned in urban planning in the past serve as a guide for future decisions.

As the availability of open data increases across all sectors, so too does the demand for transparency in decision-making.
Urban data come in different forms, they can be structured (sensors, building data, ...), semi-structured (urban system logs, ...) or unstructured (images, text, ...). Decisions are made on the basis of the data available at a given point in time. In other words, both the most recent version of the city and the previous versions are taken into account. In certain cases, complex calculations on this existing data guide the decision-makers. Reproducibility of these calculations is therefore also a requirement for transparency.

We provide the following sample queries from urban planning project proposals to better illustrate our research work:
\begin{itemize}
    \item \textbf{Q1}: \emph{Which city versions have a metro station accessible to people with disabilities?}
    \item \textbf{Q2}: \emph{Across multiple concurrent city versions, what is the maximum known height of a particular building?} (aggregation)
\end{itemize}

Our previous research work proposed the use of graph formats \cite{vinasco2021towards} for the transformation and management of heterogeneous and concurrent urban data. In this work, we want to go beyond this and explore and develop a system that can query multiple versions of the graph simultaneously to answer complex queries like the ones above. This requires not only versioning of code (complex calculations) and data. It also requires efficient querying techniques across versions.

This article briefly reviews different ways to address the need for effective tools and methodologies to analyze urban development, emphasizing the importance of versioned data management.
\section{State of the Art}
Data and code evolution have been at the heart of many recent research and industrial advances.
Taken together, they make up an important part of urban knowledge evolution. 
Given their growing use, our research is particularly focused on version control systems.

\subsection{Code and Data versioning}

Versioned repositories are systems that track and manage changes to data and code over time, allowing researchers to maintain a historical record of their work and facilitate collaboration.
When these two concepts are integrated, they provide several benefits in scientific research, such as reproducibility, transparency, and collaboration.
Version control allows different deductive paths to be explored and merged, facilitating collaboration among researchers with different expertise.
Code versioning systems like GIT and SVN play a critical role in software development, enabling collaborative work, code reuse, and traceability.
There also exist some dedicated solutions for versioning data such as DVC, DagsHub, Delta Lake, Dolt, Qri, Weights and Biases, Git LFS, Comet and LakeFS.

These different approaches are not suitable for our case study, since we want to work with concurrent versions and scenarios \cite{samuel_representation_2020}. To answer even simple queries like \textbf{Q1}, current solutions require querying multiple database instances or checking out multiple version commits, which limits query response times.

\subsection{Database versioning}
A recent interesting solution called DoltHub,  an online platform and hosting service provides version control for databases. This technology allows users to create, manage, and collaborate on databases using Git-style workflows and supports branching and merging, enabling teams to work on different features or versions of a database.
However, cross-version querying of RDF data with such a solution is a challenge. Native SPARQL (SPARQL Protocol and RDF Query Language) queries are not inherently version-aware. 
Another solution called QuitStore \cite{arndt-2020-dissertation} is an RDF data versioning system that addresses the need for efficient data retrieval across different versions. 
By implementing an RDF-based approach, QuitStore allows users to track changes and revisions to their semantic data over time. 
However, these versioning systems provide little support for cross-version queries. %, i.e., they cannot query several versions simultaneously. 
Indeed, they can either query metadata on multiple versions or query data on a single version.

Temporal databases, also known as historicized databases, are specialized databases that are designed to capture and store historical snapshots of data over a while. Some advanced temporal databases allow the analysis and querying of data at different points in time from two perspectives: how the data appeared in the real world and how it evolved within the database.
However, by their nature, such databases are limited to a linear history and cannot be directly used to store a dataset with a branching history.

\section{Contributions perspectives}

Our motivation is to find a method for retrieving knowledge from a set of urban data versions stored in RDF format\cite{vinasco2021towards}. Resource Description Framework (RDF) offers a flexible and standardized format for representing the state of the city. RDF's graph-based structure allows the integration of diverse data sources, enabling a comprehensive view of the city's attributes and relationships. By versioning the city dataset, we can systematically track and record changes, modifications, and additions over time. This comprehensive version of history provides a foundation for analyzing the city's evolution, identifying trends, and extracting valuable knowledge. For example, if we have an urban dataset, we can identify the following problem: \emph{How to analyze a set of versions of a city to produce additional knowledge?}

From a semantics point of view, a versioned graph can be assimilated to a collection of graphs, that is one graph for each version.
Together with the \texttt{GRAPH} statement in SPARQL, this provides a way to query multiple versions at the same time.
For example, the accessibility status of a given metro (\textbf{Q1}) or the height of the building (\textbf{Q2}).
However separately storing each version would cost too much space and would probably lead to unefficient query processing.

% We are currently developing ways to associate the versions and the triples contained in the associated graph.
% This means that the query engine will explore this additional information for queries across versions.

Borrowing from historicized databases, one can associate version metadata to RDF triples.
However, while a tuple in historicized database can be associated with a validity time interval, the branching nature of versioning history requires a different representation.
Using provenance techniques \cite{sikos_provenance-aware_2020} 
, this information could then be used at the query engine level to compute partial answers for several versions at once.
We aim at implementing such a query engine and compare its efficiency with solutions using existing approaches with version checkout.
We also aim at comparing the efficiency of different representations of version metadata associated to triples (for example representing the set of versions in extension or by a set of version intervals).

Note that this representation is independent from version metadata, we can thus reuse representation of version graph such as in~\cite{arndt-2020-dissertation} to trace the origin and lineage of data, for example to reference the code used to produce the data. It helps to understand its authenticity and assess its trustworthiness.

% To begin answering this question, we are exploring the concept of atomic graphs. This concept entails representing urban data as discrete, interconnected entities. This representation captures the granular changes that occur during the evolution of the city and serves as a foundation for versioning and analyzing the dataset. This modelization has been used in many research mainly to track knowledge evolution. Understanding and tracking atomic changes within the dataset allows for fine-grained versioning. By identifying and recording additions and removals, we can represent updates accurately and derive valuable insights from the evolving urban knowledge. Effective representation of updates is essential for comprehensive versioning. Techniques were used for capturing additions and removals in the dataset, enabling a clear understanding of the changes that occur over time.

% By associating metadata with each named graph, RDF-named graphs can be used to capture provenance information \cite{bizer2005ng4j}. You can include additional triples within the named graph that describe the author, the timestamp, the source, or any other relevant information about the origin and lineage of the data. This makes it possible to trace the lineage of data. It helps to understand its authenticity and assess its trustworthiness.

\begin{acks}
This work \emph{Knowledge Hub for Evolving Urban Cities} is supported and funded by the IADoc@UDL (Université de Lyon, Université de Lyon 1) and LIRIS UMR 5205. We also acknowledge the BD team and the Virtual City Project\footnote{\url{https://projet.liris.cnrs.fr/vcity/}} members for their invaluable advice and assistance.
\end{acks}

%%
%% The next two lines define the bibliography style to be used, and
%% the bibliography file.
\bibliographystyle{ACM-Reference-Format}
\bibliography{bibliography.bib}

%%
%% If your work has an appendix, this is the place to put it.
\appendix

\end{document}